# Strange Particle Production from Quark Matter Droplets

Klaus WERNER [§‡] and Michael HLADIK

*Institut für Theoretische Physik, Universität Heidelberg, Germany*

**Abstract.** We recently introduced new methods to study ultrarelativistic nuclear scattering by providing a link between the string model approach and a thermal description. The string model is used to provide information about fluctuations in energy density. Regions of high energy density are considered to be quark matter droplets and treated macroscopically. At SPS energies, we find mainly medium size droplets — with energies up to few tens of GeV. A key issue is the microcanonical treatment of individual quark matter droplets. Each droplet hadronizes instantaneously according to the available n-body phase space. Due to the huge number of possible hadron configurations, special Monte Carlo techniques have been developed to calculate this disintegration. We present results concerning the production of strange particles from such a hadronization as compared to string decay.

## I. INTRODUCTION

Studying nuclear collisions at ultrarelativistic energies ($E_{\text{cms}}$/nucleon $\gg$ 1 GeV) is motivated mainly by the expectation that a thermalized system of quarks and gluons (quark–gluon plasma) is created [1]. There are essentially two directions for modelling such interactions: dynamical and thermal approaches. The former ones refer to string models [2–7] or related methods [8], supplemented by semihard interactions at very high energies [9–12]. Here, a well established treatment of hadron-hadron scattering, based on Pomerons and AGK rules [13], is extended to nuclear interactions. Thermal methods [14–19] amount to assuming thermalization after some initial time $\tau_0$, with evolution and hadronization being mostly based on ideal gas assumptions.

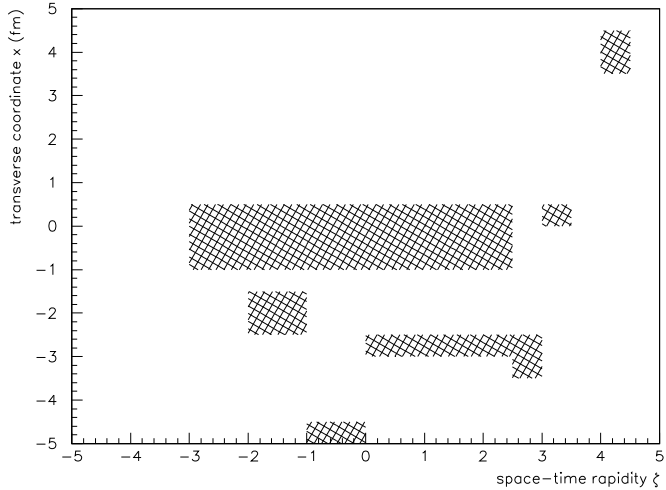

FIG. 1. High density regions in the x-$\zeta$ plane for a typical event.

Both methods have serious theoretical drawbacks. Even for nuclei as light as sulfur the string models produce particle densities that high that the hadrons are overlapping. So the independent string model is certainly too simplistic, and also considering secondary interactions as binary collisions among hadrons can theoretically not be justified. On the other hand it is also not realistic to consider a homogeneous plasma occupying the whole available volume, as assumed in thermodynamic models. To illustrate this, we show in fig. 1 a typical event of a string model simulation. We consider a central S+S collision at 200 GeV ($E_{\rm cms}/A \approx 10$ GeV), the transverse coordinates being $x$ and $y$, the longitudinal one (= beam direction) being $z$; it is useful to consider space-time rapidity $\zeta = 0.5 \ln(t+z)/(t-z)$ rather than $z$. In the figure, the hatched regions represent high energy density ($\varepsilon \geq \varepsilon_c = 1$ GeV) in the $x - \zeta$ plane ($y = 0$). We find a couple of intermediate size regions of high energy density, representing rest masses of few GeV up to few tens of GeV. This demonstrates that neither the independent string model is correct, since these high density regions cannot possibly be treated as independent hadrons nor the thermal approaches, since we do not have one big high density object but rather a couple of medium size objects in addition to plenty of ordinary hadrons and resonances in particular in the periphery.

In this paper we sketch our new approach [20,21], which is more realistic than the string model and more realistic than thermal treatments, providing a link between the two. We then show first results, where we compare rapidity and $m_t$ spectra from QM droplet hadronization with the corresponding distributions from string decay. Results from full nucleus–nucleus collisions will be published soon.

## II. THE NEW APPROACH

In this section, we review the basic features of our approach, which is meant to be a link between the "bare" string model and thermal treatments.

Based on the string model, we first determine connected regions of high energy density ($\varepsilon \geq \varepsilon_c$, for a given $\varepsilon_c$). These regions are referred to as quark matter (QM) droplets. For such regions, the initially produced hadrons serve only as a mean to produce the proper fluctuations in the energy density. Presently, a purely longitudinal expansion of the QM droplets is assumed. Once the energy density falls beyond $\varepsilon_c$, the droplet $D$ decays instantaneously into an $n$-hadron configuration $K = \{h_1 h_2 \ldots h_n\}$ with a probability proportional to $\Omega$, with $\Omega$ representing the microcanonical partition function of an $n$-hadron system. Due to the huge configuration space, sophisticated methods of statistical physics [22,23] have to be employed to solve the problem without further approximations.

Our hadronization scenario is referred to as "microcanonical hadron gas (MHG) scenario". It is certainly not the only one and probably not the most realistic one. However, we start with this scenario for a couple of reasons: the hadron gas scenario is a benchmark, widely used in the literature (in a simplified fashion though); the MHG scenario can be solved exactly; for massless hadrons even an analytical treatment is possible, providing very useful tests for the complicated numerical procedures. After having gained experience in the techniques to solve the MHG scenario, we plan to investigate alternatives as well. So the purpose of this paper is not so much to promote this particular scenario, but rather to show how a dynamical and a statistical treatment can be combined.

The first stage of our approach is the identification of high energy density regions, based on the string model, which is already discussed elsewhere [20]. Due to the empirically found correlation, $\bar{y} = \zeta$, between the average rapidity $\bar{y}$ and space-time rapidity $\zeta$, a hypersurface $\mathcal{H}_\tau$ of constant proper time $\tau$ may be introduced, in the central region simply defined by $t^2 - z^2 = \tau^2$. Appropriate coordinates on $\mathcal{H}_\tau$ are the space-time rapidity $\zeta = 0.5 \ln(t+z)/(t-z)$ and the transverse coordinates $x$ and $y$. After having used the string model (VENUS 5.08) to get complete information of hadron trajectories in space and time, we may now, for given $\tau$, determine energy densities on $\mathcal{H}_\tau$ and thus locate high density regions on $\mathcal{H}_\tau$ (with $\varepsilon > \varepsilon_c$), as shown in figure 1 for a typical example.

High density regions are considered as QM droplets, presently it is assumed that they expand purely longitudinally. Whenever other clusters or hadrons cross their way, the two objects fuse to form a new, more energetic cluster. Due to the expansion, the energy density of a cluster will at some stage drop below $\varepsilon_c$, which causes an instantaneous decay.

We employ the "microcanonical hadron gas (MHG) scenario" for the hadronization: the probability of a droplet $D$ — characterized by the invariant mass $E$, the volume $V$, and the flavour $Q = (Q^u, Q^d, \ldots)$ — to decay into a hadron configuration $K = \{h_1, \ldots, h_n\}$ of hadrons $h_i$ is given as

$$\text{prob}(D \to K) \sim \Omega(K) , \tag{1}$$

with $\Omega(K)$ being the microcanonical partition function of an ideal, relativistic gas of the $n$ hadrons $h_i$,

$$\Omega(K) = C_{\text{vol}} C_{\text{deg}} C_{\text{ident}} \phi , \tag{2}$$

with

$$C_{\text{vol}} = \frac{V^n}{(2\pi)^{3n}} , \quad C_{\text{deg}} = \prod_{i=1}^{n} g_i , \quad C_{\text{ident}} = \prod_{\alpha \in \mathcal{S}} \frac{1}{n_\alpha!} . \tag{3}$$

Here, $C_{\text{deg}}$ accounts for degeneracies ($g_i$ is the degeneracy of particle $i$), and $C_{\text{ident}}$ accounts for the occurence of identical particles in $K$ ($n_\alpha$ is the number of particles of species $\alpha$). The last factor

$$\phi = \int \prod_{i=1}^{n} d^3 p_i \, \delta(E - \Sigma \varepsilon_i) \, \delta(\Sigma \vec{p}_i) \, \delta_{Q, \Sigma q_i} \tag{4}$$

is the so-called phase space integral, with $\varepsilon_i = \sqrt{m_i^2 + p_i^2}$ being the energy and $\vec{p}_i$ the 3-momentum of particle $i$. The vector $q_i = (q_i^u, q_i^d, \ldots)$ represents the flavour content of hadron $i$. The expression eq. (4) is valid for the centre-of-mass frame of the droplet $D$.

We have to define a set $\mathcal{S}$ of hadron species; we include the pseudoscalar and vector mesons $(\pi, K, \eta, \eta', \rho, K^*, \omega, \phi)$ and the lowest spin-$\frac{1}{2}$ and spin-$\frac{3}{2}$ baryons $(N, \Lambda, \Sigma, \Xi, \Delta, \Sigma^*, \Xi^*, \Omega)$ and the corresponding antibaryons. A configuration is then an arbitrary set $\{h_1, \ldots, h_n\}$ with $h_i \in \mathcal{S}$.

We are interested in droplet masses from few GeV up to $10^3$ GeV, corresponding to particle numbers $n = |K|$ between 2 and $10^3$. So we have to deal with a huge configuration space, which requires to employ Monte Carlo techniques, well known in statistical physics. The method at hand is to construct a Markov process, specified by an initial configuration $K_0$, and a transition probability matrix $p(K_t \to K_{t+1})$. In generating a sequence $K_0, K_1, K_2, \ldots$, two fundamental issues have to be payed attention at:

- initial transient: starting usually off equilibrium, it takes a number of iterations, $I_{\text{eq}}$, before one reaches equilibrium;

- autocorrelation in equilibrium: even in equilibrium, subsequent configurations, $K_a$ and $K_{a+i}$, are correlated for some range $I_{\text{auto}}$ of $i$.

In general, both $I_{\text{eq}}$ and $I_{\text{auto}}$ should be as small as possible.

We are going to proceed as follows: for a given droplet $D$ with mass $E$, volume $V$, and flavour $Q$, we start from some initial configuration $K_0$, and generate a sequence $K_0, K_1, \ldots, K_{I_{\text{eq}}}$, with $I_{\text{eq}}$ being sufficiently large to have reached equilibrium. If we repeat this procedure many times, getting configurations $K_{I_{\text{eq}}}^{(1)}, K_{I_{\text{eq}}}^{(2)}, \ldots$, these configurations are distributed as $\Omega(K)$. So for our problem, we have only to deal with the initial transient, not with the autocorrelation in equilibrium. We have to find a transition probability $p$ such that it leads to an equilibrium distribution $\Omega(K)$, with the initial transient $I_{\text{eq}}$ being as small as possible.

Sufficient for the appropriate convergence to $\Omega(K)$ is the detailed balance condition,

$$\Omega(K_a)\, p(K_a \to K_b) = \Omega(K_b)\, p(K_b \to K_a) \;, \tag{5}$$

and ergodicity, which means that for any $K_a, K_b$ there must exist some $r$ with the probability to get in $r$ steps from $K_a$ to $K_b$ being nonzero. Henceforth, we use the abbreviations

$$\Omega_a := \Omega(K_a); \quad p_{ab} := p(K_a \to K_b). \tag{6}$$

Following Metropolis [22], we make the ansatz

$$p_{ab} = w_{ab}\, u_{ab} \;, \tag{7}$$

with a so-called proposal matrix $w$ and an acceptance matrix $u$. Detailed balance now reads

$$\frac{u_{ab}}{u_{ba}} = \frac{\Omega_b}{\Omega_a} \frac{w_{ba}}{w_{ab}} \;, \tag{8}$$

which is obviously fulfilled for

$$u_{ab} = F\left(\frac{\Omega_b}{\Omega_a} \frac{w_{ba}}{w_{ab}}\right) \;, \tag{9}$$

with some function $F$ fulfilling $F(z)\,/\,F(z^{-1}) = z$. Following Metropolis [22], we take

$$F(z) = \min(z, 1) \;. \tag{10}$$

The power of the method is due to the fact that an arbitrary $w$ may be chosen, in connection with $u$ being given by eq. (9). So our task is twofold: we have to develop an efficient algorithm to calculate, for given $K$, the weight $\Omega(K)$, and we have to find an appropriate proposal matrix $w$ which leads to fast convergence (small $I_{\text{eq}}$). The first task can be solved, a detailed description will be published soon [24]. In the following we discuss about constructing an appropriate matrix $w$.

Most natural, though not necessary, is to consider symmetric proposal matrices, $w_{ab} = w_{ba}$, which simplifies the acceptance matrix to $u_{ab} = F(\Omega_b/\Omega_a)$. This is usually referred to as Metropolis algorithm. Whereas for spin systems, it is obvious how to define a symmetric matrix $w$, this is not so clear in our case. We may take spin systems as guidance. A configuration $K$ is per def. a set of hadrons $\{h_1, \ldots, h_n\}$ with the ordering not being relevant, so $\{\pi^0, \pi^0, p\}$ is the same as $\{p, \pi^0, \pi^0\}$. We introduce "microconfigurations" to be sequences $\{h_1, \ldots, h_n\}$ of hadrons, where the ordering does matter. So for a given configuration $K_a = \{h_1, \ldots, h_n\}$ there exist several microconfigurations $\tilde{K}_{aj} = \{h_{\pi_j(1)}, \ldots, h_{\pi_j(n)}\}$, with $\pi_j$ representing a permutation. The weight of a microconfiguration is

$$\Omega(\tilde{K}_{aj}) = \frac{1}{n!} \left\{ \prod_{\alpha \in \mathcal{S}} n_\alpha! \right\} \Omega(K_a) , \qquad (11)$$

with $n_\alpha$ being the number of hadrons of type $\alpha$. Taking for example $K = \{p, \pi^0, \pi^0\}$, there are three microconfigurations $\{p, \pi^0, \pi^0\}$, $\{\pi^0, p, \pi^0\}$ and $\{\pi^0, \pi^0, p\}$, with weight $\Omega(K)/3$.

So far we deal with sequences $\{h_1, \ldots, h_n\}$ of arbitrary length $n$, to be compared with spin systems with fixed lattice size. We therefore introduce "zeros", i.e. we supplement the sequences $\{h_1, \ldots, h_n\}$ by adding $L-n$ zeros, as $\{h_1, \ldots, h_n, 0, \ldots, 0\}$, to obtain sequences of fixed length $L$. The zeros may be inserted at any place, not necessarily at the end. Therefore the weight of a microconfiguration $K_{aj}$ with zeros relative to the one without, $\tilde{K}_{aj}$, is one divided by the number of possibilities to insert $L-n$ zeros, so from eq. (11) we get

$$\Omega(K_{aj}) = \frac{1}{n!} \left\{ \prod_{\alpha \in \mathcal{S}} n_\alpha! \right\} \frac{n!\,(L-n)!}{L!} \Omega(K_a) . \qquad (12)$$

We now have the analogy with a spin system: we have a one-dimensional lattice of fixed size $L$, with each lattice site containing either a hadron or a zero. Henceforth, we use for microconfigurations with zeros the notation $K_{aj} = \{h_1, \ldots, h_L\}$ with $h_i$ being a hadron or zero.

Since from now on we only consider microconfigurations with zeros ($K_{aj}$) rather than configurations ($K_a$), we are going to write $K_a$ instead of $K_{aj}$, keeping in mind that $a$ represents a double index, and say "configuration" rather than "microconfiguration with zeros". The advantage is that we can use the above formulas specifying the Metropolis algorithm without changes.

We are now in a position to define a symmetric proposal matrix $w(K_a \to K_b)$, with $K_a = \{h_1^a, \ldots, h_L^a\}$ and $K_b = \{h_1^b, \ldots, h_L^b\}$, as

$$w(K_a \to K_b) = \frac{2}{L(L-1)} \left\{ \prod_{\substack{k=1 \\ k \neq i,j}}^{L} \delta_{h_k^a h_k^b} \right\} v(h_i^a h_j^a \to h_i^b h_j^b) , \qquad (13)$$

with

$$v(h_i^a h_j^a \to h_i^b h_j^b) = \begin{cases} |\mathcal{P}(h_i^a h_j^a)|^{-1} & \text{if } h_i^b h_j^b \in \mathcal{P}(h_i^a h_j^a) \\ 0 & \text{else} \end{cases} , \qquad (14)$$

where $\mathcal{P}(h_i^a h_j^a)$ is the set of all pairs $(h_i h_j)$ with the same total flavour as the pair $(h_i^a h_j^a)$. The symbol $|\mathcal{P}|$ refers to the number of pairs of $\mathcal{P}$. The term $\{\}$ in eq. (13) makes sure that up to one pair all hadrons in $K_a$ and $K_b$ are the same, the term $2/L(L-1)$ is the probability to randomly choose some pair of lattice indices $i$ and $j$. So our proposal matrix amounts to randomly choosing a pair in $K_a$, and replacing this pair by some pair with the same flavour, with all possible replacements having the same weight. The proposal matrix is obviously symmetric, since $v$ is symmetric (the symmetry of $v$ is crucial!). We have now fully defined an algorithm, which due to general theorems will converge, but how fast, i.e., how large is $I_{\mathrm{eq}}$? Considering particle ratios, like $n_{\pi^0}/n_{\pi^+}$, we find immediately that we have a very slow convergence, so $I_{\mathrm{eq}}$ is too large for the method to be of practical importance. This is obvious, since the method is not very democratic: flavourless particles like $\pi^0, \rho^0$ or also zeros are much more frequently proposed than all the rest. This shortcoming can be fixed by defining $w$ such that two pairs are exchanged rather than one, the first pair being replaced by a completely arbitrary pair, the second one by some pair to guarantee flavour conservation. In addition it is necessary to weight the "zeros" differently than the hadrons. This improved method violates the symmetry of $w$, however, the asymmetry $w_{ab}/w_{ba}$ can be calculated and properly taken into account. Further details of the "asymmetric algorithm" will be published elsewhere [24].

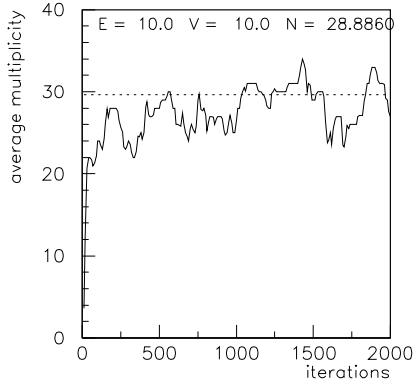

FIG. 2. Multiplicity versus the number of iterations.

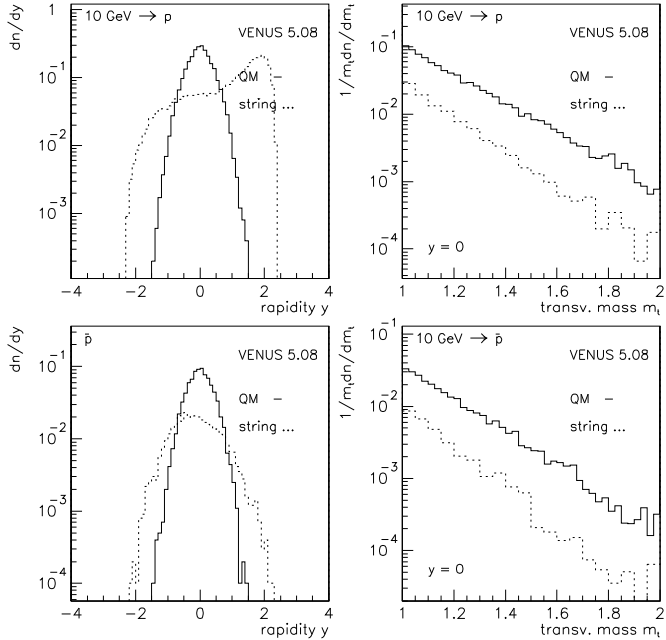

FIG. 3. Comparing hadronization of quark matter droplets (solid) and string decay (dotted): Rapidity spectra (left) and transverse mass distributions (right) of protons (top) and antiprotons (bottom).

The asymmetric method converges quite fast. As a check, we consider massless hadrons, where analytical results exist. In fig. 2 we plot the iterated total multiplicity $N$ for a droplet with $E = 10$ GeV and $V = 10$ fm$^3$, compared to the average multiplicity $\bar{N}$ from the analytical calculation (dashed line). The "equilibration time" $I_{\text{eq}}$ is roughly given as

$$I_{\text{eq}} / \bar{N} \approx 10 ,  \qquad (15)$$

with the initial configuration being two $\pi_0$'s.

## III. RESULTS

In the following we compare hadron production from QM droplets — as discussed in the previous section — with string decay. We consider droplets with $E = 10$ GeV, $V = 10$ fm$^3$ and $uud$-flavour and correspondingly $u$-$ud$ strings with an energy of 10 GeV. We are going to present rapidity ($y$) and transverse mass ($m_t$) spectra for different hadrons, in particular strange ones. In fig. 3, we show results for protons and antiprotons. Here — as in the following figures — the solid histograms represent QM droplets, the dashed ones strings. For protons from strings, we clearly observe a "leading particle effect", i.e. a maximum of the $y$ distribution at forward rapidities. This is due to the fact that the fast forward $ud$ diquark is constituent of the observed proton. In case of antiprotons, we have a "real" production process with no string (or droplet) constituents appearing in the hadron. We clearly observe more antiprotons from QM droplets compared to string decay, the $m_t$ spectra in the QM case are somewhat flatter. The (anti)lambda spectra, shown in fig. 4, are quite similar to the (anti)proton case, but the $\bar{\Lambda}$ enhancement (of QM compared to string decay) is large, roughly a factor of ten. For the (anti)cascades, presented in fig. 5, we find even more enhancement, about a factor of fifty.

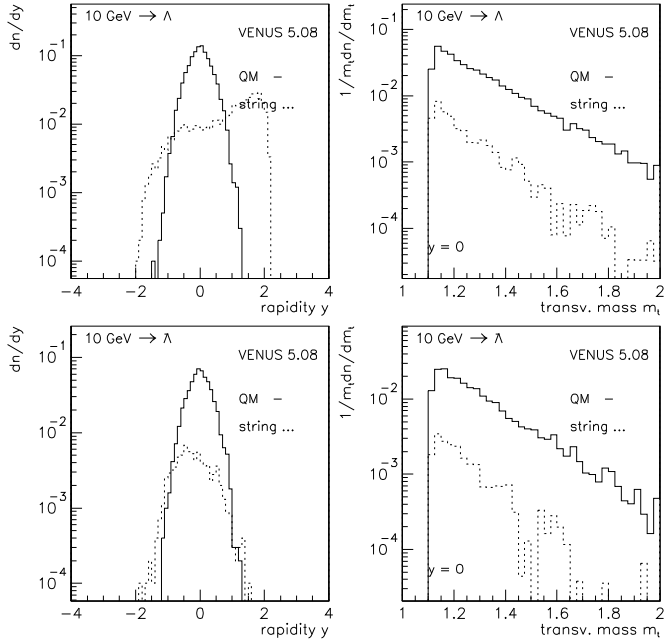

FIG. 4. Comparing hadronization of quark matter droplets (solid) and string decay (dotted): Rapidity spectra (left) and transverse mass distributions (right) of lambdas (top) and antilambdas (bottom).

The baryon–antibaryon ($B\bar{B}$) production may be summarized as follows: whereas $B\bar{B}$ pairs are rarely produced from string decay, QM droplets provide considerably larger rates. The enhancement increases with increasing strangeness content (the string results drop dramatically). Such large enhancement (of 10 or 50) will not show up in nucleus–nucleus collisions, since here we have a mixture of hadrons from droplets and those from string decay (formed in dilute areas).

We now turn to mesons. The pion rapidity spectra from strings, shown in fig. 6, are much broader than the spectra from droplet decay, the absolute numbers (integrals) being not too different though. Also the $m_t$ spectra are quite similar. For the kaons ($K^+$), as shown in fig. 7, we find some enhancement, much less though as compared with $\bar{\Lambda}$ and $\bar{\Xi}$.

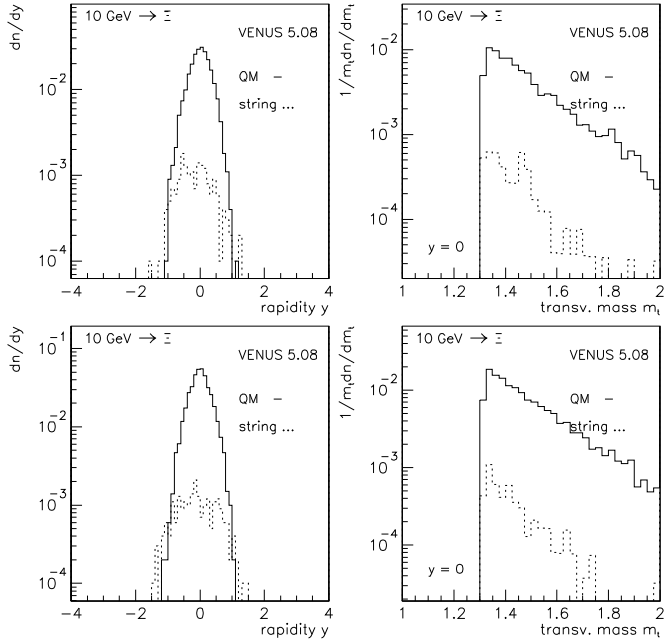

FIG. 5. Comparing hadronization of quark matter droplets (solid) and string decay (dotted): Rapidity spectra (left) and transverse mass distributions (right) of cascades (top) and anticascades (bottom).

So comparing droplet with string decay, we find a substantial enhancement of antibaryons — the more strangeness the better. This is due to the fact that $B\bar{B}$ production is strongly suppressed in string decay, with no such suppression acting in droplet decay.

## IV. SUMMARY

We have presented our new method to model ultrarelativistic scattering, where one attempts to link the string model and statistical approaches. The string model is used to locate high density regions, which are then referred to as quark-matter droplets and treated macroscopically. We presented first results which showed that antibaryons, in particular strange ones, are much more frequent in droplet decay as compared to string decay. This will lead to an enhanced antibaryon rate in nuclear collisions as compared to proton-proton.

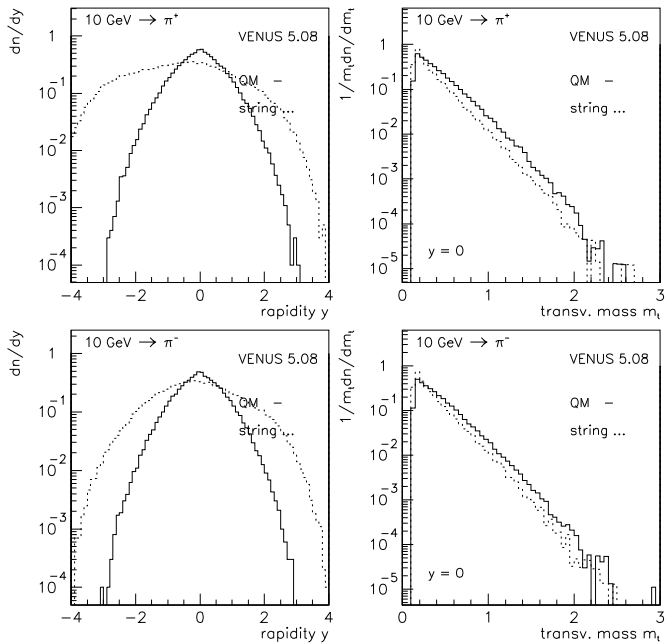

FIG. 6. Comparing hadronization of quark matter droplets (solid) and string decay (dotted): Rapidity spectra (left) and transverse mass distributions (right) of $\pi^+$'s (top) and $\pi^-$'s (bottom).

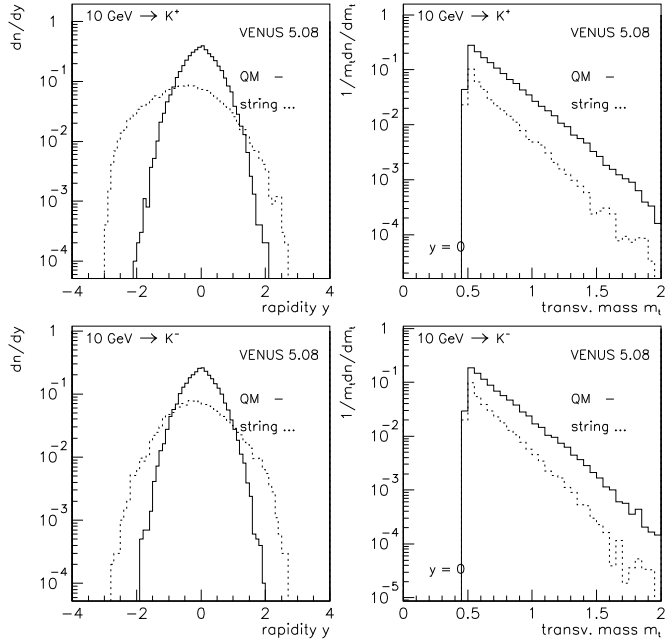

FIG. 7. Comparing hadronization of quark matter droplets (solid) and string decay (dotted): Rapidity spectra (left) and transverse mass distributions (right) of $K^+$'s (top) and $K^-$'s (bottom).